\documentclass[preprint,showpacs,preprintnumbers,amsmath,amssymb]{revtex4}

\usepackage{graphicx}% Include figure files
\usepackage{dcolumn}% Align table columns on decimal point
\usepackage{bm}% bold math

\begin{document}

%\preprint{APS/123-QED}

\title{Quantum key distribution with dual detectors}% Force line breaks with \\

\author{Bing Qi}

\author{Yi Zhao}

\author{Xiongfeng Ma}

\author{Hoi-Kwong Lo}

\author{Li Qian}

\affiliation{ Center for Quantum Information and Quantum Control
(CQIQC),
Dept. of Physics and Dept. of Electrical and Computer Engineering,\\
University of Toronto, Toronto, M5S 3G4, Canada }

%\date{\today}% It is always \today, today,
             %  but any date may be explicitly specified

\begin{abstract}
To improve the performance of a quantum key distribution (QKD)
system, high speed, low dark count single photon detectors (or low
noise homodyne detectors) are required. However, in practice, a fast
detector is usually noisy. Here, we propose a ``dual detectors''
method to improve the performance of a practical QKD system with
realistic detectors: the legitimate receiver randomly uses either a
fast (but noisy) detector or a quiet (but slow) detector to measure
the incoming quantum signals. The measurement results from the quiet
detector can be used to bound eavesdropper's information, while the
measurement results from the fast detector are used to generate
secure key. We apply this idea to various QKD protocols. Simulation
results demonstrate significant improvements in both BB84 protocol
with ideal single photon source and Gaussian-modulated coherent
states (GMCS) protocol; while for decoy-state BB84 protocol with
weak coherent source, the improvement is moderate. We also discuss
various practical issues in implementing the ``dual detectors''
scheme.
\end{abstract}

% PACS, the Physics and Astronomy
                             % Classification Scheme.
%\keywords{Suggested keywords}%Use showkeys class option if keyword
                              %display desired
\maketitle

\section{Introduction}

One important practical application of quantum information is
quantum key distribution (QKD), whose unconditional security is
based on the fundamental laws of quantum mechanics
\cite{no_cloning,BB84,E91,CV,nature2003,securityproof,ILD01,GLLP}.
In principle, any eavesdropping attempts by a third party (Eve) will
unavoidably introduce quantum bit errors. So, it is possible for the
legitimate users (Alice and Bob) to upper bound the amount of
information acquired by the eavesdropper given system parameters and
the measured quantum bit error rate (QBER). If the QBER is not too
high and the transmission efficiency is not too low, Alice and Bob
can then distill a final secure key by performing error correction
(to correct errors due to imperfections in the QKD system and errors
due to eavesdropping) and privacy amplification (to remove Eve's
information on the final key).

A practical QKD system has imperfections, which will contribute to
QBER even in the absence of Eve. If Alice and Bob cannot distinguish
the intrinsic QBER due to imperfections from the one induced by Eve,
in order to guarantee the unconditional security, they have to
assume that all errors originate from eavesdropping. Under this
assumption, the intrinsic QBER will increase the costs for both
error correction and privacy amplification. On the other hand, if
Alice and Bob do have a way to distinguish the intrinsic QBER from
the one due to eavesdropping, then the cost for the privacy
amplification can be reduced \cite{Gisin}.

One important error source in a practical QKD system is the noise of
the receiver's detector, for example, the dark count probability of
a single photon detector (SPD) or the ``excess noise" of a homodyne
detector. As the distance between Alice and Bob increases (which is
equivalent to a higher channel loss), the contribution to QBER from
detector's noise becomes more significant. When the QBER is over some
threshold, no secure key can be generated. The maximum secure
distance of a QKD system is thus limited by the detector's noise. On
the other hand, the secure key rate is proportional to the operating
rate of the QKD system, which is mainly determined by the speed of
the detector. In brief, an ideal detector should be fast and
noiseless. Unfortunately, in practice, high speed detectors are
usually noisy.

In classical metrology, there are many elegant methods to combat
various noises associated with the measurement devices. It is
natural to ask this question: can we introduce classical
``calibration" processes into a QKD system to deal with various
noises associated with its intrinsic imperfections? An intuitive
idea is as follow: the receiver, Bob, adds a high speed optical
switch at the entrance of his device. He uses this switch to
randomly block some input signals. The measurement results with no
input signal can be used to estimate the intrinsic noise of the
detector. Alice and Bob can further estimate among the total QBER
(measured when Bob's switch is open), how much is contributed by
this intrinsic detector noise. The QBER caused by the intrinsic
detector noise does not contribute to Eve's information, only the
QBER above it does. Since Alice and Bob can bound Eve's information
more tightly, the cost for privacy amplification will be lowered. We
remark that the cost for error correction remains the same, because
whether the error is caused by eavesdropping or by the intrinsic
noise, Alice and Bob will treat them equally during an error
correction process.

Note that there is an implicit assumption in the above argument:
that Eve cannot control the intrinsic noise of the detector, or at
most, she can increase but not decrease it. If Eve can decrease the
detector noise when the switch is ON, the above argument is not
valid because Bob cannot use the detector noise measured with switch
OFF to estimate the detector noise with switch ON. Unfortunately,
this assumption is not straightforward to justify. The first rule in
quantum cryptography is: to guarantee unconditional secure, one
should make assumptions that are most favorable to Eve. In this
case, we allow Eve to fully control the noise of Bob's detectors,
and thus the above intuitive idea does not work.

Here we propose a ``dual detectors'' method to improve the
performance of a QKD system based on realistic detectors. The basic
idea is quite simple: Bob has two detectors, one is fast but noisy,
while the other one is quiet but slow. For each incoming quantum
signal, Bob randomly chooses to use either the fast detector (with a
high probability) or the slow detector (with a low probability) to
do the measurement. During the classical data post-processing stage,
Alice and Bob use the QBER measured by the slow (quiet) detector to
bound Eve's information, and they use the raw key bits from the fast
detector to produce a secure key. Since Eve cannot predict which
detector Bob will choose for each individual bit, her attack is
independent on which detector is used. So, Alice and Bob can apply
the bound (about Eve's information) acquired from the low-noise
detector to the raw key acquired from the fast (but noisy) detector.
By using a tighter bound on Eve's information, the cost for privacy
amplification will be reduced. Intuitively, our proposal will
improve the performance of practical QKD setups.

In this paper, we apply the ``dual detectors" idea into three
different protocols: namely, the BB84 protocol with perfect single
photon source \cite{BB84}(Section II), the decoy state BB84 protocol
with weak coherent source
\cite{decoy_theory,Ma05,decoy_experiment,TES2}(Section III), and the
Gaussian-modulated coherent states (GMCS) protocol \cite{nature2003}
(Section IV). Our simulation results confirmed the intuitive
prediction of performance, demonstrating significant improvements in
both BB84 protocol with an ideal single photon source and GMCS
protocol; while for decoy-state BB84 protocol with a weak coherent
state source, the improvement is moderate. In Section V, we discuss
some practical issues in the implementation of the ``dual
detectors'' idea, including the loss introduced by the optical
switch, the distribution of the signals between two detectors, the
dispersion of a long fiber and the security of a practical setup.
Finally, in Section VI, we end this paper with a brief discussion on
the security of a practical QKD system.

\section{Single photon BB84 QKD with dual detectors}

The most well known and mature QKD protocol is BB84 protocol
\cite{BB84}. There have been a lot of research activities in
building a practical single photon source \cite{singlephotonsource}.
In this section, we assume that an ideal single photon source is
employed. In this case, the secure key rate is given by \cite{GLLP}
\begin{equation}
R=\frac{1}{2}rQ_{1}[1-f(e_{1})H_{2}(e_{1})-H_{2}(e_{1})].
\end{equation}
Here the factor $1/2$ is due to half of the time, Alice and Bob use
different bases (if one uses the efficient BB84 protocol
\cite{efficient_QKD}, this factors is one). $r$ is the pulse
repetition rate of the QKD system. $Q_1$ is the overall gain (taking
into account of channel loss, optical loss inside Bob and the
detection efficiency of SPD), which is defined as the ratio of Bob's
detection events to the total signal pulses sent by Alice. $e_1$ is
the QBER. $f(x)$ is the bidirectional error correction efficiency,
and $H_{2}(x)$ is the binary entropy function, given by
\begin{equation}
H_{2}(x)=-x\log_{2}(x)-(1-x)\log_{2}(1-x).
\end{equation}
Note that in Eq.(1), the term $f(e_{1})H_{2}(e_{1})$ is the cost for
error correction, while the term $H_{2}(e_{1})$ is the cost for
privacy amplification. With ``dual detectors" method, Alice and Bob
use a ``quiet" SPD (which yields a lower QBER at a long distance) to
give a tighter bound on Eve's information $H_{2}(e_{1})$. This
tighter bound can be used to lower the cost of privacy
amplification when Alice/Bob use a ``noisier" (but faster) SPD to
generate the secure key.

Note the ``dual detectors'' method cannot be simply explained as
using the quiet detector to estimate the dark count of the noisy
detector. It should be understood as using the quiet detector to
bound Eve's information more tightly. For each pulse from Alice,
right beyond Bob's optical switch (for randomly choosing detector),
Eve's potential information, $I_{\mathrm{Eve}}^{(0)}$, is
independent on which detector Bob will choose to do the measurement.
We can imagine Bob's two detectors as two independent QKD systems.
Either of them can upper bound Eve's information properly. This
means the two bounds (on Eve's information) acquired from the two
detectors satisfy: $I_{\mathrm{Eve}}^{(1)}\geq
I_{\mathrm{Eve}}^{(0)}$, and $I_{\mathrm{Eve}}^{(2)}\geq
I_{\mathrm{Eve}}^{(0)}$. So, Bob can use either of
$I_{\mathrm{Eve}}^{(1)}$ (which is quantified by the QBER measured
with detector 1) or $I_{\mathrm{Eve}}^{(2)}$ to perform the privacy
amplification without compromising the security of the system.

We model the QKD system as follows \cite{Ma05}. The gain of the QKD
system is
\begin{equation}
Q_{1}=Y_{0}+G_{\mathrm{ch}}G_{\mathrm{Bob}}\eta_\mathrm{D}
\end{equation}
where $Y_{0}$ is the background rate, $G_{\mathrm{ch}}$ is the
channel transmission efficiency, $G_{\mathrm{Bob}}$ is the optical
transmittance in Bob's system, and $\eta_\mathrm{D}$ is the
efficiency of the SPD. Here we assume that $Y_{0}\ll1$ and
$G_{\mathrm{ch}}G_{\mathrm{Bob}}\eta_\mathrm{D}\ll1$. The quantum
channel between Alice and Bob is telecom fiber with attenuation
$\alpha=0.21\mathrm{dB/km}$. The channel efficiency can be estimated
by $G_{\mathrm{ch}}=10^{-\alpha L/10}$, where $L$ is the fiber
length in km.

The QBER is determined by
\begin{equation}
e_{1}=\frac{e_{0}Y_{0}+e_{\mathrm{det}}G_{\mathrm{ch}}G_{\mathrm{Bob}}\eta_\mathrm{D}}{Q_{1}}
\end{equation}
Here $e_{0}=0.5$ is the error rate of background counts, which is
dominated by dark counts \cite{Dark_count}, and $e_{\mathrm{det}}$
is the probability that a single photon hits the wrong detector when
Alice and Bob choose the same basis. $e_{\mathrm{det}}$
characterizes the alignment and the stability of the optical system
and the cross-talk between adjacent signals, etc.

We assume that Bob randomly chooses to use one of the following two
SPDs: the first one is fast but noisy (with operating rate $r_1$,
efficiency $\eta_{\mathrm{D}}^{(1)}$ and dark count probability
$Y_{0}^{(1)}$), while the second one is slow but quiet (with
operating rate $r_2$, efficiency $\eta_{\mathrm{D}}^{(2)}$ and dark
count probability $Y_{0}^{(2)}$ ). To improve the overall efficiency
(only the fast SPD contributes to the final secure key), the
probability of choosing the slow SPD should be small (in asymptotic
case, it can approach zero). The secure key rate of the ``dual
detectors" scheme is given by
\begin{equation}
R=\frac{1}{2}r_{1}Q_{1}^{(1)}[1-f(e_{1}^{(1)})H_{2}(e_{1}^{(1)})-H_{2}(e_{1}^{(2)})].
\end{equation}
Here, $e_{1}^{(1)}$ and $e_{1}^{(2)}$ are the QBERs measured by SPD1
and SPD2, respectively.

Numerical simulations have been performed based on different
combinations of SPDs.

\subsection{Case One: up-conversion SPD and transition-edge sensor SPD}

Two different types of SPD are employed in this case. SPD1 is a high
speed SPD based on up-conversion process. Recently, these MHz
devices have been employed in GHz rate QKD systems
\cite{upconversion1,upconversion2}. SPD2 is a ``low noise'' SPD
based on transition-edge sensors (TESs) \cite{TES1,TES2}. Simulation
parameters are summarized as follows: $\alpha=0.21\mathrm{dB/km}$,
$f(x)=1.22$; $G_{\mathrm{Bob}}=0.16$ and $e_{\mathrm{det}}=0.018$
\cite{TES1}; $r_1=1\mathrm{GHz}$, $\eta_{\mathrm{D}}^{(1)}=0.059$
and $Y_{0}^{(1)}=1.3\times10^{-5}$ \cite{upconversion2};
$r_2=2.5\mathrm{MHz}$, $\eta_{\mathrm{D}}^{(2)}=0.5$ and
$Y_{0}^{(2)}=3\times10^{-7}$\cite{TES2}.

Fig.1 shows the simulation results. The key rate of
``dual-detector'' system is higher than either of the two single SPD
systems up to $\sim124\mathrm{km}$. Note that, at a long distance,
the system with SPD2 alone yields a higher key rate than a
dual-detector system. Thus Bob can simply use SPD2 alone.

\begin{figure}[!t]
\resizebox{8.5cm}{!}{\includegraphics{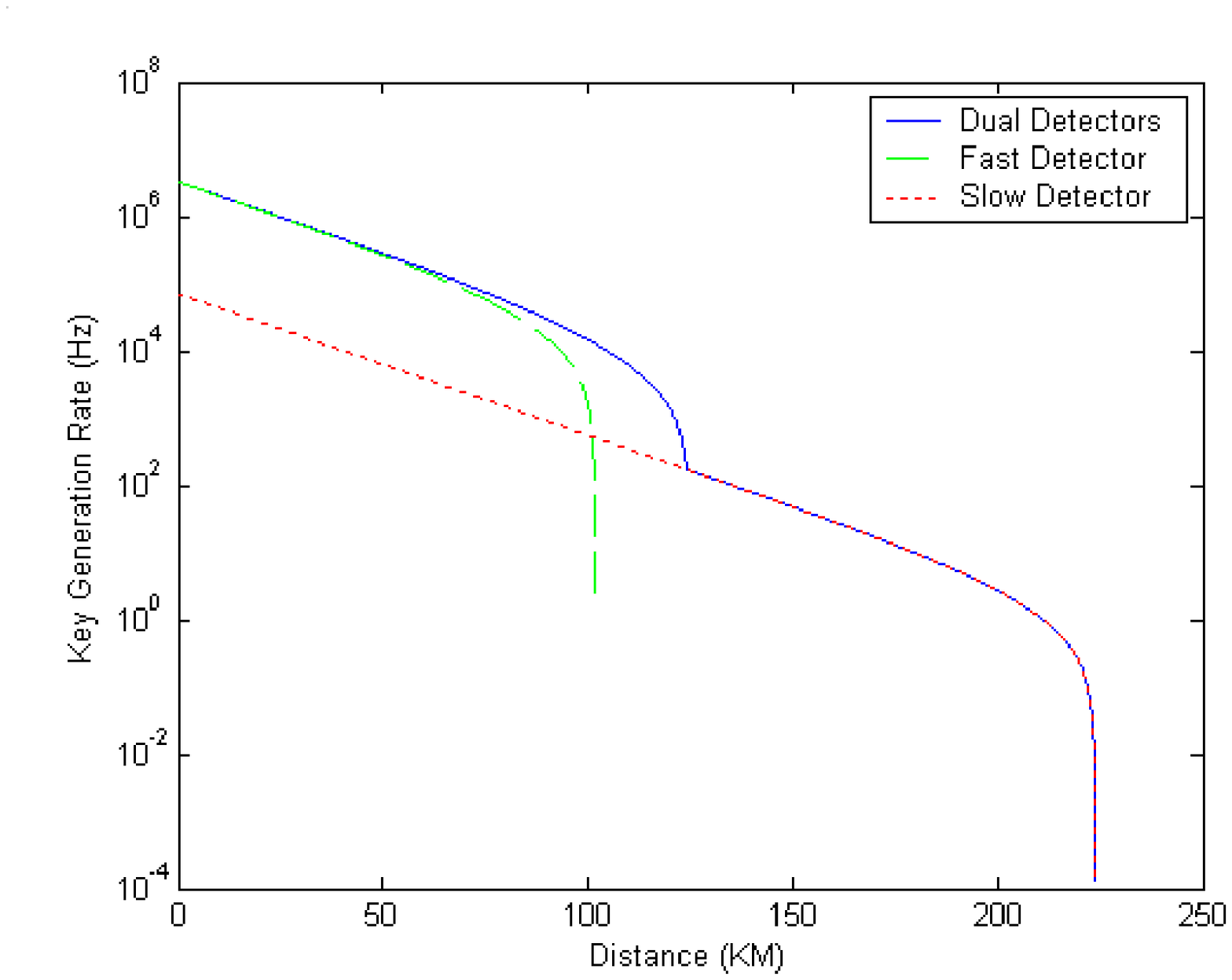}}
\caption{Simulation results for BB84 protocol with single photon
source. Simulation parameters: $\alpha=0.21\mathrm{dB/km}$,
$f(x)=1.22$; $G_{\mathrm{Bob}}=0.16$ and $e_{\mathrm{det}}=0.018$
\cite{TES1}; $r_1=1\mathrm{GHz}$, $\eta_{\mathrm{D}}^{(1)}=0.059$
and $Y_{0}^{(1)}=1.3\times10^{-5}$ \cite{upconversion2};
$r_2=2.5\mathrm{MHz}$, $\eta_{\mathrm{D}}^{(2)}=0.5$ and
$Y_{0}^{(2)}=3\times10^{-7}$\cite{TES2}. The key rate of ``dual
detectors" system is higher than either of the two single SPD
systems up to $\sim124\mathrm{km}$. Note that at a longer distance,
when the system with SPD2 alone yields a higher key rate, Bob can
simply use SPD2 itself.}\label{fig1}
\end{figure}

\subsection{Case Two: low jitter up-conversion SPD and transition-edge sensor SPD}

In this case, we assume that SPD1 is a low jitter up-conversion
SPD\cite{SPD_10G1}, which has been applied in a $10\mathrm{GHz}$ QKD
system \cite{SPD_10G2}. Note that, in this case, due to the high
pulse repetition rate and non-zero time jitter, the cross-talk
between adjacent pulses is high. This contributes to a high QBER
independent of fiber length, which is equivalent to a high
$e_{\mathrm{det}}$ for SPD1. The parameters for SPD1 are:
$r_1=10\mathrm{GHz}$, $\eta_{D}^{(1)}=0.0027$,
$Y_{0}^{(1)}=3.2\times10^{-9}$ and $e_{\mathrm{det}}^{(1)}=0.097$
\cite{SPD_10G2}. Other parameters are the same as in Case One.

Fig.2 shows the simulation results. The key rate of the ``dual
detectors'' system is significantly higher than either of the two
single SPD systems up to $\sim200\mathrm{km}$. Here we particularly
remark that no secure key can be produced by SPD1 alone at any
distance.

\begin{figure}[!t]
\resizebox{8.5cm}{!}{\includegraphics{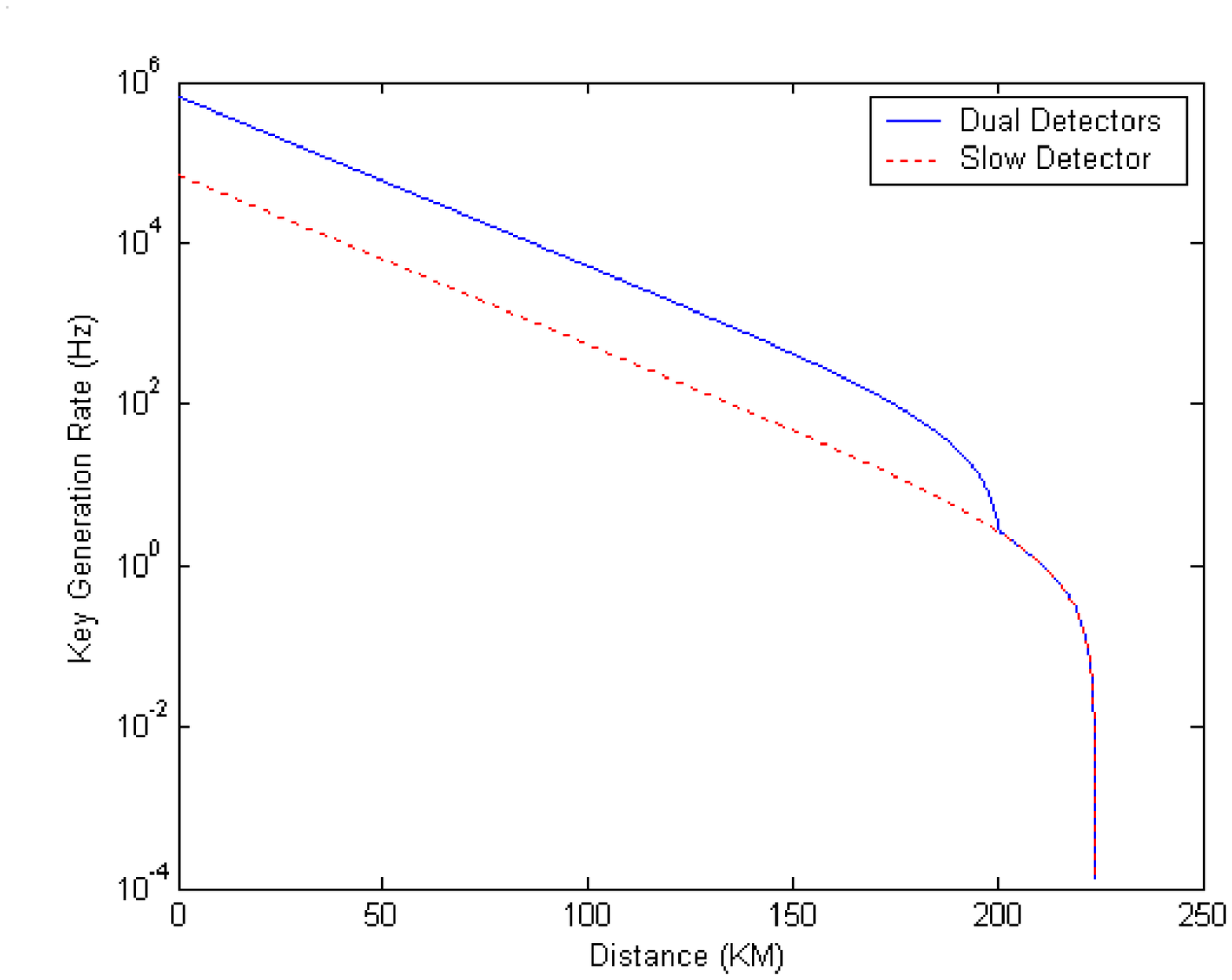}}
\caption{Simulation results for BB84 protocol with single photon
source. Simulation parameters: $r_1=10\mathrm{GHz}$,
$\eta_{\mathrm{D}}^{(1)}=0.0027$, $Y_{0}^{(1)}=3.2\times10^{-9}$ and
$e_{\mathrm{det}}^{(1)}=0.097$ \cite{SPD_10G2}. Other parameters are
same as in Fig.1. The key rate of ``dual detectors" system is
significantly higher than either of the two single SPD systems up to
$\sim200\mathrm{km}$. Note that no secure key can be produced by
SPD1 alone at any distance.}\label{fig2}
\end{figure}

\subsection{Case Three: two low jitter up-conversion SPDs}

In Case One and Case Two, the working principles of the two SPDs are
substantially different. To prevent Eve from exploring the
difference between the two detectors, special counter measures, such
as narrowband filters, may be required. We will discuss this topic
in details in Section V.

In Case Three, two identical low jitter SPDs are employed to remove
the asymmetry between the two detectors. The probability for
choosing SPD1 is close to one. So, it still suffers from the high
QBER due to the cross-talk between adjacent pulses. Since the
probability for choosing SPD2 is quite small (say $<0.01$), the
cross-talk between adjacent pulses can be neglected, and the QBER
from SPD2 will be much lower. Simulation parameters are summarized
as follows: $r_1=10\mathrm{GHz}$, $\eta_{\mathrm{D}}^{(1)}=0.0027$,
$Y_{0}^{(1)}=3.2\times10^{-9}$ and $e_{\mathrm{det}}^{(1)}=0.097$
\cite{SPD_10G2}. $r_2=100\mathrm{MHz}$,
$\eta_{\mathrm{D}}^{(2)}=0.0027$, $Y_{0}^{(2)}=3.2\times10^{-9}$ and
$e_{\mathrm{det}}^{(2)}=0.018$. Other parameters are the same as
before.

Fig.3 shows the simulation results. The key rate of the ``dual
detectors" system is significantly higher than either of the two
single SPD systems up to $\sim190\mathrm{km}$. Again, no secure key
can be produced by SPD1 alone at any distance.

\begin{figure}[!t]
\resizebox{8.5cm}{!}{\includegraphics{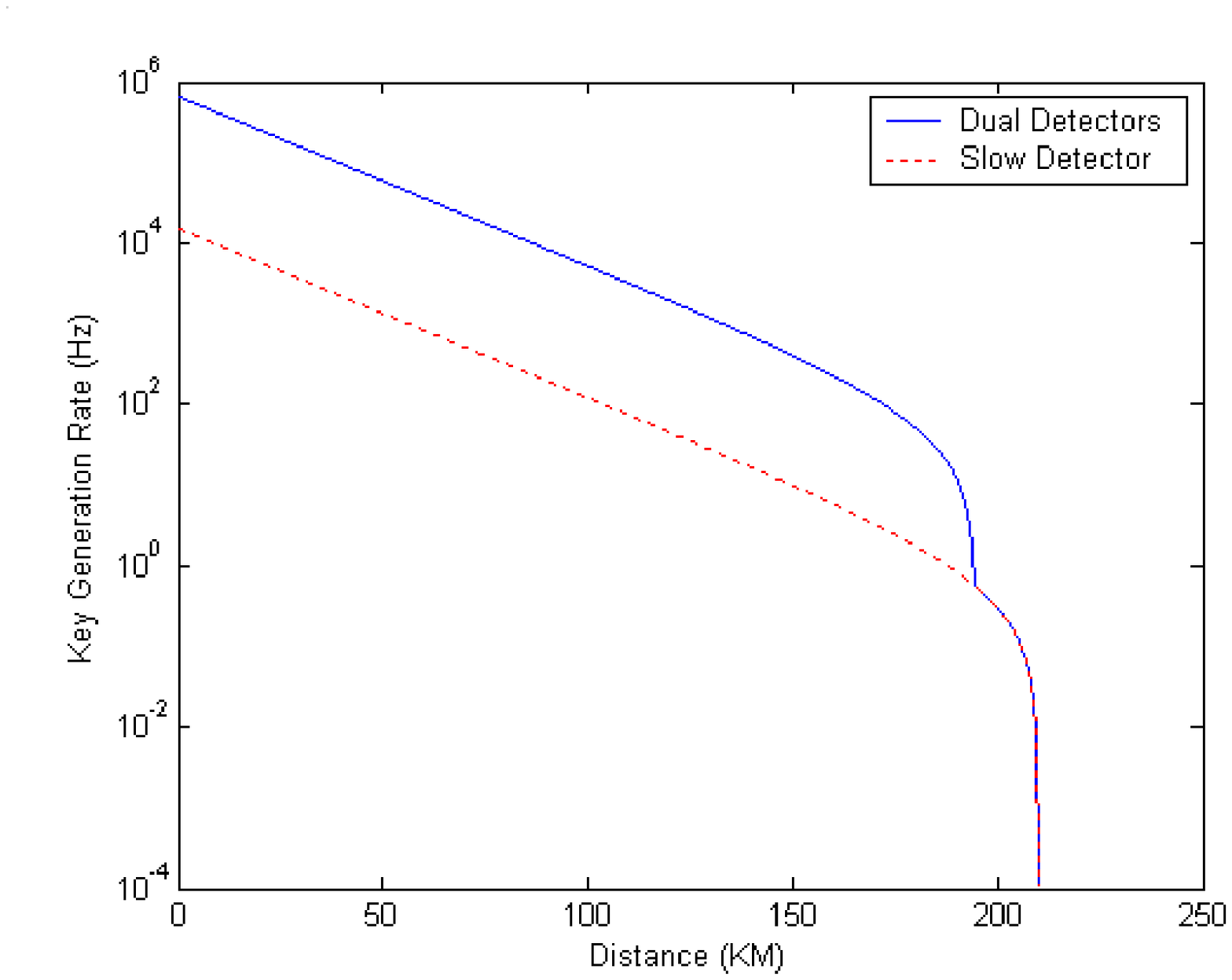}}
\caption{Simulation results for BB84 protocol with single photon
source. Simulation parameters: $r_1=10\mathrm{GHz}$,
$\eta_{\mathrm{D}}^{(1)}=0.0027$, $Y_{0}^{(1)}=3.2\times10^{-9}$ and
$e_{\mathrm{det}}^{(1)}=0.097$ \cite{SPD_10G2};
$r_2=100\mathrm{MHz}$, $\eta_{\mathrm{D}}^{(2)}=0.0027$,
$Y_{0}^{(2)}=3.2\times10^{-9}$ and $e_{\mathrm{det}}^{(2)}=0.018$;
Other parameters are same as in Fig.1. The key rate of ``dual
detectors" system is significantly higher than either of the two
single SPD systems up to $\sim190\mathrm{km}$. Note that no secure
key can be produced by SPD1 alone at any distance.}\label{fig3}
\end{figure}

In summary, our simulation results demonstrate that the ``dual
detectors'' method can improve the performance of single photon BB84
QKD system dramatically. We remark that the same idea can also be
applied to QKD with imperfect single photon sources.

\section{Decoy state BB84 QKD with dual detectors}

Currently, most of QKD experiments are performed with a weak
coherent source. The photon number of each pulse follows a Poisson
distribution with a parameter $\mu$ as its expected photon number,
which is set by Alice. In this case, the secure key rate is given by
\cite{GLLP}
\begin{equation}
R=\frac{1}{2}r[Q_{1}-f(E_{\mu})Q_{\mu}H_{2}(E_{\mu})-Q_{1}H_{2}(e_{1})].
\end{equation}
Here $Q_{\mu}$, $E_{\mu}$ are the gain and the overall QBER of
signal states, while $Q_{1}$, $e_{1}$ are the gain and the QBER of
single-photon components. Note that only $Q_{\mu}$, $E_{\mu}$ can be
determined from experimental data directly, while the bounds on
$Q_{1}$ and $e_{1}$ have to be estimated from the specific QKD
protocol and model of QKD system.

Here, we assume that Alice and Bob perform ideal decoy state BB84
protocol \cite{decoy_theory,Ma05}. In the asymptotic case, the
estimated value of the above four parameters are given by
\cite{Ma05}
\begin{align}
%\begin{equation}
Q_{\mu}=Y_{0}+1-e^{-\eta\mu}\\
E_{\mu}=[e_{0}Y_{0}+e_{\mathrm{det}}(1-e^{-\eta\mu})]/Q_{\mu}\\
Q_{1}=(Y_{0}+\eta)\mu e^{-\mu}\\
e_{1}=(e_{0}Y_{0}+e_{\mathrm{det}}\eta)\mu e^{-\mu}/Q_{1}
%\end{equation}
\end{align}
Here $\eta=G_{\mathrm{ch}}G_{\mathrm{Bob}}\eta_{\mathrm{D}}$ is the
overall efficiency of the QKD system.

The optimal $\mu$ for the signal state can be estimated from \cite{Ma05}
\begin{equation}
(1-\mu)e^{-\mu}=\frac{f(E_{\mu})H_{2}(e_{\mathrm{det}})}{1-H_{2}(e_{\mathrm{det}})}
\end{equation}

With the ``dual detectors'' method, we expect that Alice and Bob can
obtain a tighter bound on $e_{1}$, thus lowering the cost of privacy
amplification. Simulation parameters are summarized as follows:
$\alpha=0.21\mathrm{dB/km}$, $f(x)=1.22$, $\mu=0.73$ ;
$G_{\mathrm{Bob}}=0.16$ and $e_{\mathrm{det}}=0.018$ \cite{TES1};
$r_1=1\mathrm{GHz}$, $\eta_{\mathrm{D}}^{(1)}=0.059$ and
$Y_{0}^{(1)}=1.3\times10^{-5}$ \cite{upconversion2};
$r_2=2.5\mathrm{MHz}$, $\eta_{\mathrm{D}}^{(2)}=0.5$ and
$Y_{0}^{(2)}=3\times10^{-7}$ \cite{TES2}. The optimal $\mu$ in the
case of ``dual detectors'' is chosen based on the parameters of the
fast detector. The simulation results are shown in Fig.4. We see
moderate improvement up to $\sim82\mathrm{km}$.

\begin{figure}[!t]
\resizebox{8.5cm}{!}{\includegraphics{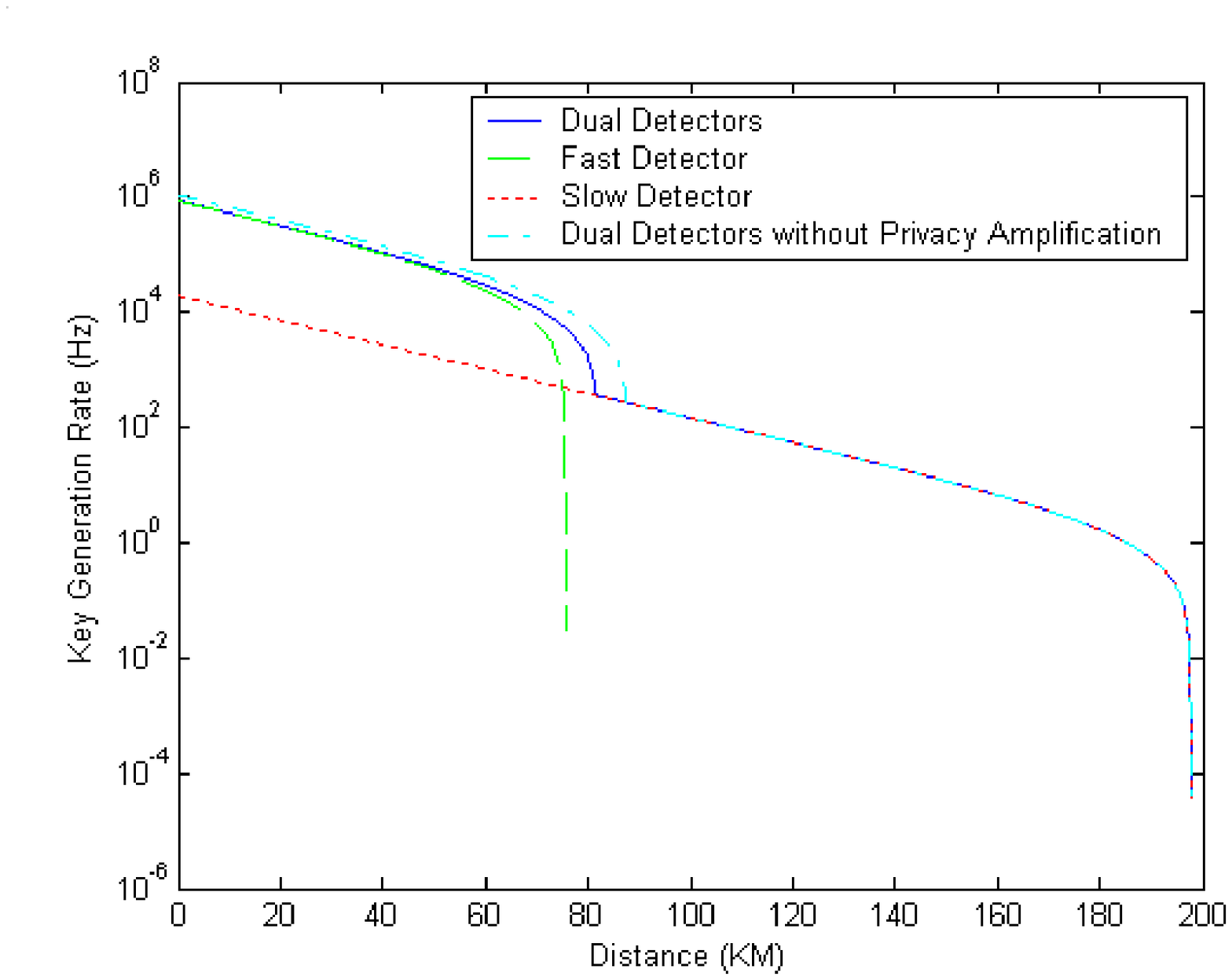}}
\caption{Simulation results for Decoy state BB84 protocol with weak
coherent source: Simulation parameters: $\alpha=0.21\mathrm{dB/km}$,
$f(x)=1.22$, $\mu=0.73$; $G_{\mathrm{Bob}}=0.16$ and
$e_{\mathrm{det}}=0.018$ \cite{TES1}; $r_1=1\mathrm{GHz}$,
$\eta_{\mathrm{D}}^{(1)}=0.059$ and $Y_{0}^{(1)}=1.3\times10^{-5}$
\cite{upconversion2}; $r_2=2.5\mathrm{MHz}$,
$\eta_{\mathrm{D}}^{(2)}=0.5$ and
$Y_{0}^{(2)}=3\times10^{-7}$\cite{TES2}.The key rate of ``dual
detectors'' system is higher than either of the two single SPD
systems up to $\sim82\mathrm{km}$. Note that even without doing any
privacy amplification, the improvement in this case is
moderate.}\label{fig4}
\end{figure}

The limited improvement in this protocol can be understood from
Eq.(6). The second term ($f(E_{\mu})Q_{\mu}H_{2}(E_{\mu})$) at the
right hand side of Eq.(6) is the cost for error correction, while
the third term ($Q_{1}H_{2}(e_{1})$) is the cost for privacy
amplification. Since $f(E_{\mu})Q_{\mu}H_{2}(E_{\mu})$ is
significantly larger than $Q_{1}H_{2}(e_{1})$, the cost of the error
correction term is the dominating factor. The ``dual detectors''
system only allows us to reduce the privacy amplification term, but
not the error correction term. Therefore, any improvement due to the
``dual detectors'' system for decoy state BB84 protocol over telecom
fibers will be moderate. This point is clearly illustrated by our
numerical simulations in Fig.4: even if Alice and Bob did not
perform any privacy amplification, the improvements in secure key
rate and secure distance would still be moderate.

\section{Gaussian-modulated coherent states QKD with dual detectors}

Recently, GMCS QKD has drawn a lot of attention for its potential
high secure key rate, especially at relatively short distance
\cite{nature2003,dr,rr,GMCS,PP}. In this protocol \cite{nature2003},
Alice draws two random numbers $X_A$ and $P_A$ from a Gaussian
distribution with mean zero and variance $V_A$ (in shot-noise
units), and sends a coherent state $|X_A+iP_A\rangle$ to Bob. Bob
randomly chooses to measure either the phase quadrature or the
amplitude quadrature with a phase modulator and a homodyne detector.
During the classical communication stage, Bob informs Alice which
quadrature he measures for each pulse and Alice will drop the other
one. Eventually, they can work out a set of correlated Gaussian
variables, which will be converted to a secure key. It has been
shown in \cite{nature2003} that with ``reverse reconciliation'' (RR)
protocol \cite{rr}, this scheme can tolerate high channel loss on
the condition that the excess noise (the noise above vacuum noise)
is not too high, while with ``direct reconciliation'' (DR) protocol
\cite{dr}, this scheme can yield a high key rate at relatively short
distances.

\subsection{Direct Reconciliation Protocol}

We assume symmetry on the noise characteristics between the
amplitude quadrature measurement and phase quadrature measurement.
For additive Gaussian noise channels, the mutual information between
Alice and Bob, $I_{AB}$, and between Alice and Eve, $I_{AE}$, are
given by \cite{dr}
\begin{align}
%\begin{equation}
I_{AB} =(1/2)\log_2 [(V+\chi)/(1+\chi)]\\
%\end{equation}
%\begin{equation}
I_{AE} =(1/2)\log_2 [(V+1/\chi)/(1+1/\chi)]
%\end{equation}
\end{align}
where $V=V_A+1$ is the variance of Alice's field quadratures in
shot-noise units, $\chi=\chi_{\mathrm{vac}}+\varepsilon$ is the
equivalent input noise, where $\chi_{\mathrm{vac}}=(1-G)/G$ is the
``vacuum noise'' associated with the overall transmission efficiency
$G$, while $\varepsilon$ is the ``excess noise''.
$G=G_{\mathrm{ch}}G_{\mathrm{det}}$, where $G_{\mathrm{ch}}$ is the
channel efficiency and $G_{\mathrm{det}}$ is the detection
efficiency.

Note that since $\varepsilon$ is the ``excess noise'' with respect
to the input, it can be described by
$\varepsilon=\varepsilon_{\mathrm{pre}}+\varepsilon_{\mathrm{det}}/G$,
where $\varepsilon_{\mathrm{pre}}$ and $\varepsilon_{\mathrm{det}}$
are the ``excess noises'' associated with imperfections in state
preparation and homodyne detection, respectively. Obviously, at long
distances (i.e., $G$ is small), the main contribution to
$\varepsilon$ is from the detector noise.

The security key rate of a DR protocol is given by \cite{dr}
\begin{equation}
R_1=r(\beta I_{AB}-I_{AE})
\end{equation}
where $r$ is the repetition rate of the QKD system and
$\beta\in(0,1)$ is the efficiency of DR protocol.

In GMCS QKD system, the ``excess noise'' plays a similar role as the
dark count probability of SPD in BB84 protocol. The ``dual
detectors'' scheme can be employed to improve the performance of a
GMCS QKD system based on realistic homodyne detectors, as in the
case of BB84 protocol. Specifically, at the classical communication
stage, Alice and Bob use the measurement results from the quiet
detector and Eq.(13) to estimate $I_{AE}$ and the measurement
results from the fast detector and Eq.(12) to calculate $I_{AB}$.
Using Eqs.(12-14), the secure key rate of the ``dual detectors''
scheme can be derived as
\begin{align}
%\begin{equation}
R_2=r_{1}\{(\beta/2)\log_2[(V+\chi_{\mathrm{vac}}+\varepsilon_{1})/(1+\chi_{\mathrm{vac}}+\varepsilon_{1})]\nonumber \\
-(1/2)\log_2[(V+1/(\chi_{\mathrm{vac}}+\varepsilon_{2}))/(1+1/(\chi_{\mathrm{vac}}+\varepsilon_{2}))]\}
%\end{equation}
\end{align}

Simulation parameters are summarized as follows:
$\alpha=0.21\mathrm{dB/km}$, $V=40$, $\beta=1$,
$G_{\mathrm{det}}=0.80$;
$\varepsilon_{\mathrm{pre}}=0.05$\cite{GMCS};
$r_{1}=82\mathrm{MHz}$, $\varepsilon_{\mathrm{det1}}=0.43$
\cite{fasthom}; $r_{2}=1\mathrm{MHz}$,
$\varepsilon_{\mathrm{det2}}=0.01$\cite{GMCS}.

Fig.5 shows the simulation results. With the ``dual detectors''
method, we see a significant improvement of the key rate (more than
one order) at relatively short distance (up to $5$km).

\begin{figure}[!t]
\resizebox{8.5cm}{!}{\includegraphics{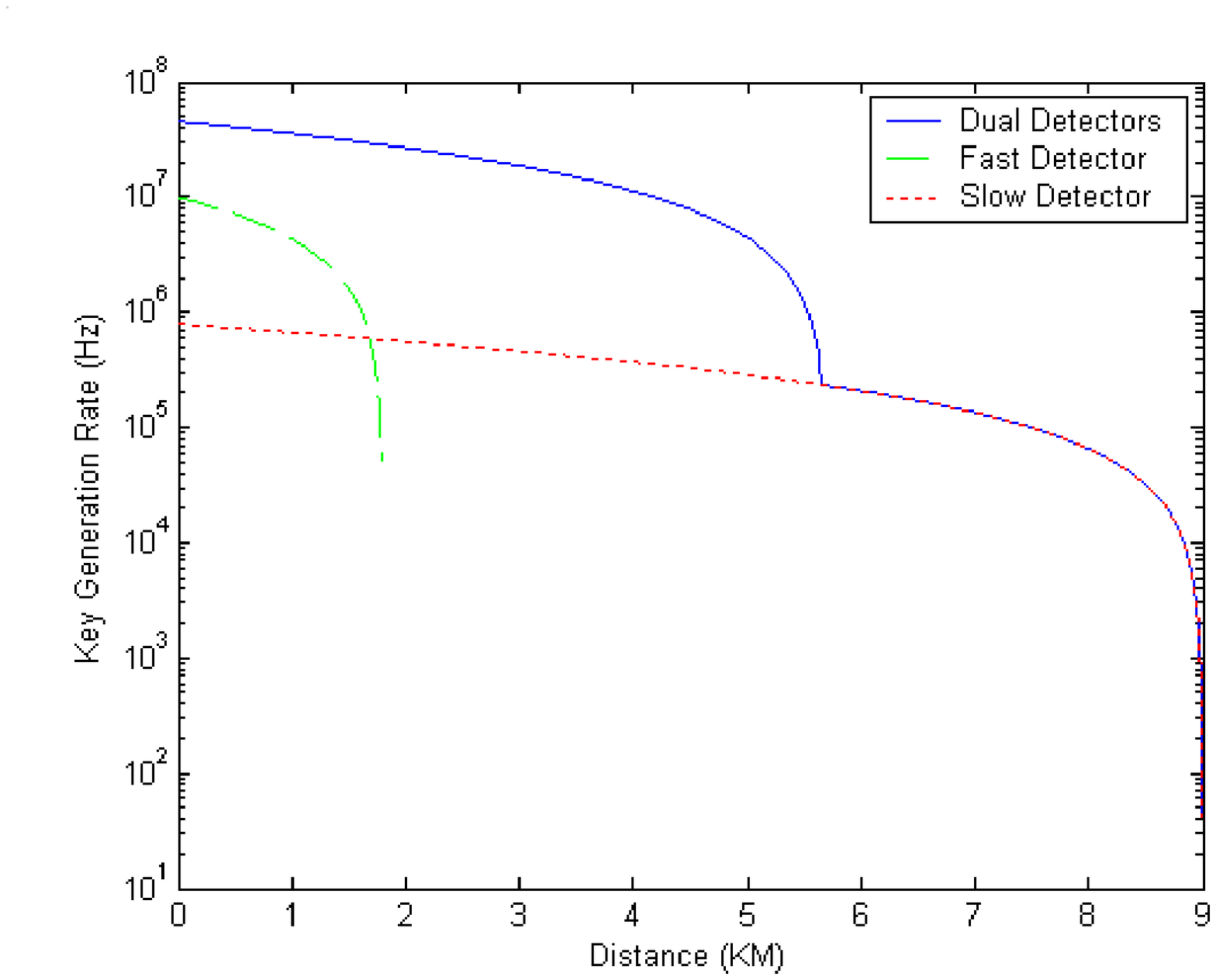}}
\caption{Simulation results for GMCS QKD with DR protocol.
Simulation parameters: $\alpha=0.21\mathrm{dB/km}$, $V=40$,
$\beta=1$; $G_{\mathrm{det}}=0.8$,
$\varepsilon_{\mathrm{pre}}=0.05$\cite{GMCS};
$r_{1}=82\mathrm{MHz}$, $\varepsilon_{\mathrm{det1}}=0.43$
\cite{fasthom}; $r_{2}=1\mathrm{MHz}$,
$\varepsilon_{\mathrm{det2}}=0.01$\cite{GMCS}. With the
``dual-detectors'' method, we see a significant improvement of the
key rate (more than one order of magnitude) at relatively short
distance(up to $5$km).}\label{fig5}
\end{figure}

\subsection{Reverse Reconciliation Protocol}

In RR protocols, Bob sends classical information to Alice, who in
turn modifies her initial data to match with Bob's measurement
results. The security key rate of a RR protocol is given by
\cite{nature2003,GMCS}
\begin{equation}
R_1=r(\beta I_{BA}-I_{BE})
\end{equation}
where the mutual information between Bob and Alice, $I_{BA}$, and
between Bob and Eve, $I_{BE}$, are given by \cite{nature2003}
\begin{align}
%\begin{equation}
I_{BA} =(1/2)\log_2 [(V+\chi)/(1+\chi)]\\
%\end{equation}
%\begin{equation}
I_{BE} =(1/2)\log_2[G^{2}(V+\chi)(V^{-1}+\chi)]
%\end{equation}
\end{align}
We remark that to derive the above equations, Eve is allowed to
control both the efficiency and excess noise in Bob's system. In
contrast, in \cite{nature2003,GMCS}, the authors took a
``realistic'' approach by assuming that the noises associated with
Bob's system do not contribute to Eve's information.

We remark that there is a substantial difference between GMCS QKD
with DR protocol and GMCS QKD with RR protocol. In DR protocol,
Alice/Bob try to bound the mutual information between Alice and Eve
$I_{AE}^{(0)}$, which is independent on the performance of Bob's
measurement device. Due to the noise and loss presented in Bob's
system, they will overestimate $I_{AE}^{(0)}$ as $I_{AE}^{(1)}$
(with detector1) or $I_{AE}^{(2)}$ (with detector2). Obviously, they
can use $\mathrm{min}\{I_{AE}^{(1)}, I_{AE}^{(2)}\}$ as an
estimation of $I_{AE}^{(0)}$ in Eq.(14). In ``reverse
reconciliation'' method, the above argument cannot be applied. In
this case, Alice/Bob try to bound the mutual information between Bob
and Eve $I_{BE}$, which depends on both the efficiency and the noise
of the homodyne detector (see Eq.(18), where the overall
transmission efficiency $G$ contains contribution from the
efficiency of the homodyne detector ). If the efficiencies of the
two detectors are different, Eve's information on Bob's measurement
results acquired with detector1 may be different from her
information on Bob's measurement results acquired with detector2. In
order to use the slow detector to give a better bound on $I_{BE}$
for the data acquired with the fast detector, we have to assume that
both detectors have the same efficiency. Note that this is a
reasonable assumption in practice, since the efficiency of the
homodyne detector is mainly determined by two factors---the optical
coupling efficiency and the quantum efficiency of the photo diode.
Both factors are insensitive to the operation rate.

We remark that transmission loss plays different roles in different
QKD protocols. In GMCS QKD, the transmission loss will introduce
``vacuum noise'' to Bob's measurement results, and Bob cannot
distinguish this ``vacuum noise'' from the ``excess noise''
contributed by the homodyne detector or other imperfections in the
QKD system. To bound Eve's information on Bob's measurement results,
Alice and Bob have to estimate both the efficiency of the QKD system
and the ``excess noise''. On the other hand, in BB84 QKD, since
Alice/Bob post-select the cases when Bob has detections (they drop
all the other cases), the transmission loss only lower the
efficiency but not contribute to the QBER. To bound Eve's
information, Alice/Bob only need to estimate the QBER. This may
explain why in BB84 QKD, to apply the ``dual detectors'' idea, it is
no necessary to make assumptions on the efficiencies of the two
detectors.

The secure key rate of the ``dual detectors'' scheme can be derived
as
\begin{align}
%\begin{equation}
R_2=r_{1}\{(\beta/2)\log_2[(V+\chi_{\mathrm{vac}}+\varepsilon_{1})/(1+\chi_{\mathrm{vac}}+\varepsilon_{1})]\nonumber \\
-(1/2)\log_2[G^{2}(V+\chi_{\mathrm{vac}}+\varepsilon_{2})(V^{-1}+\chi_{\mathrm{vac}}+\varepsilon_{2})]\}
%\end{equation}
\end{align}

Simulation parameters are the same as in DR protocol. Fig.6 shows
the simulation results. With the ``dual detectors'' method, we see a
significant improvement of the key rate (more than one order of
magnitude) at relatively short distance (up to $17$km).  Note that,
in this case, no positive key rate can be achieved with detector1
alone at any distance.

\begin{figure}[!t]
\resizebox{8.5cm}{!}{\includegraphics{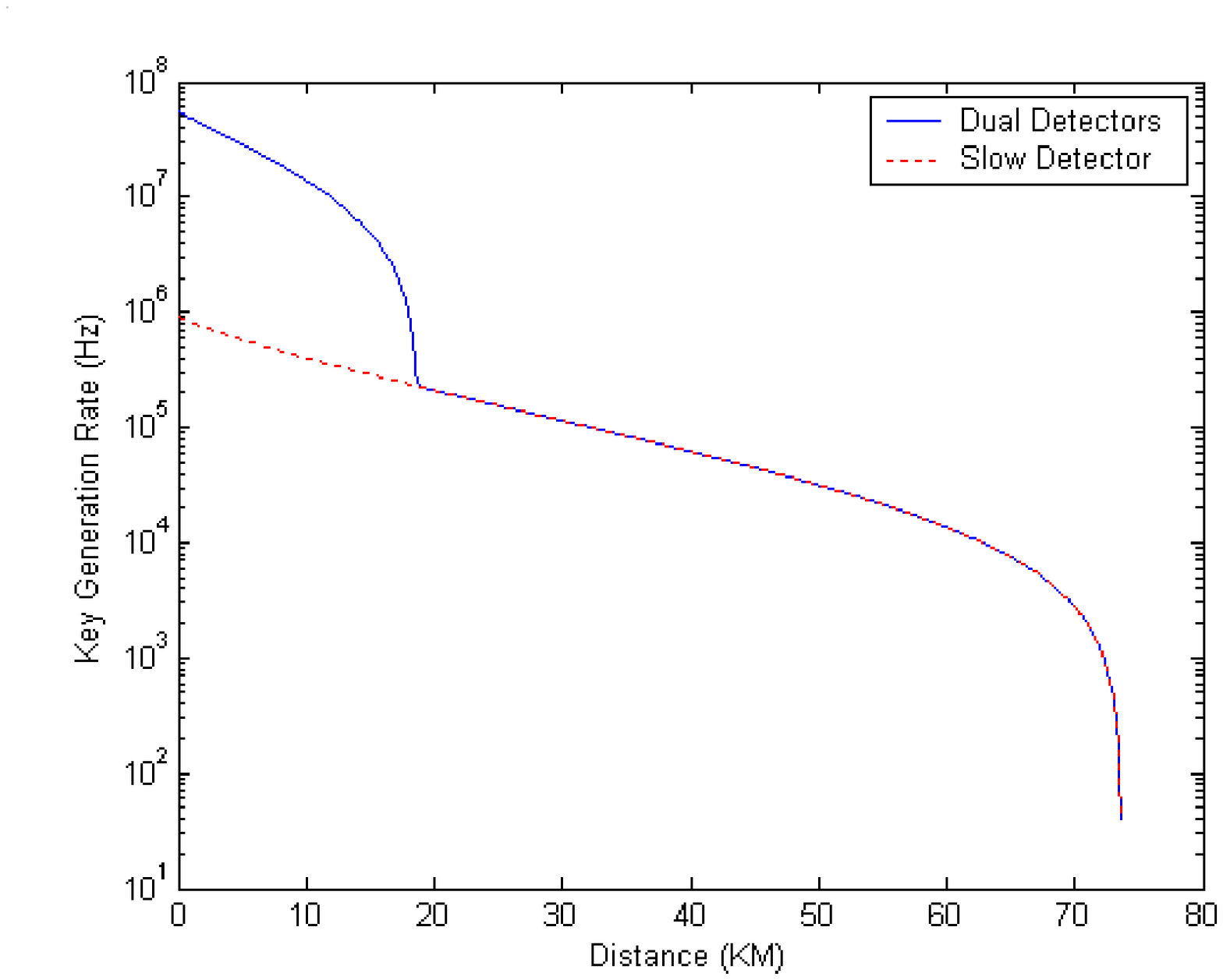}}
\caption{Simulation results for GMCS QKD with RR protocol.
Simulation parameters: $\alpha=0.21\mathrm{dB/km}$, $V=40$,
$\beta=1$; $G_{\mathrm{det}}=0.8$,
$\varepsilon_{\mathrm{pre}}=0.05$\cite{GMCS};
$r_{1}=82\mathrm{MHz}$, $\varepsilon_{\mathrm{det1}}=0.43$
\cite{fasthom}; $r_{2}=1\mathrm{MHz}$,
$\varepsilon_{\mathrm{det2}}=0.01$\cite{GMCS}. With the
``dual-detectors'' method, we see a significant improvement of the
key rate (more than one order of magnitude) at relatively short
distance(up to $17$km). Note that in this case, no positive key rate
can be achieved with detector1 alone at any distance. }\label{fig6}
\end{figure}

In practice, for a finite key length, the reconciliation algorithm
is not perfect. Fig.7 shows the simulation results with a realistic
RR protocol ($V=20$, $\beta=0.8$, other parameters are the same as
in Fig.6). With the ``dual detectors'' method, we see a significant
improvement of the key rate (more than one order of magnitude) at
relatively short distance (up to $5$km). Again, no positive key rate
can be achieved with detector1 alone at any distance.

\begin{figure}[!t]
\resizebox{8.5cm}{!}{\includegraphics{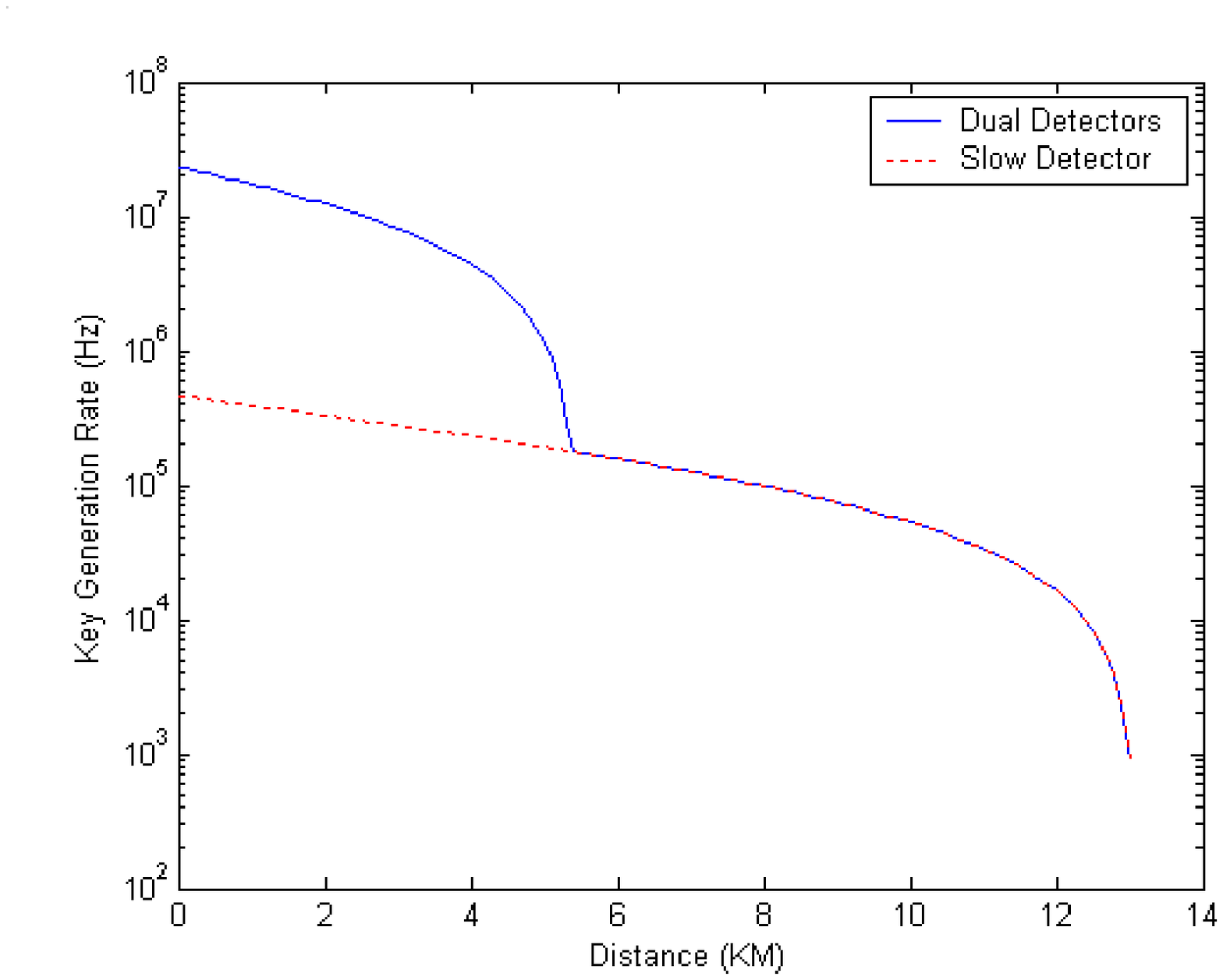}}
\caption{Simulation results for GMCS QKD with realistic RR protocol.
Simulation parameters: $\alpha=0.21\mathrm{dB/km}$, $V=20$,
$\beta=0.8$; $G_{\mathrm{det}}=0.8$,
$\varepsilon_{\mathrm{pre}}=0.05$\cite{GMCS};
$r_{1}=82\mathrm{MHz}$, $\varepsilon_{\mathrm{det1}}=0.43$
\cite{fasthom}; $r_{2}=1\mathrm{MHz}$,
$\varepsilon_{\mathrm{det2}}=0.01$\cite{GMCS}. With the
``dual-detectors'' method, we see a significant improvement of the
key rate (more than one order of magnitude) at relatively short
distance(up to $5$km). Note that in this case, no positive key rate
can be achieved with detector1 alone at any distance. }\label{fig7}
\end{figure}

We remark that the above security analysis about GMCS QKD, which are
cited from \cite{nature2003}, may be applicable to individual
attacks only. The security of GMCS protocol under the most general
attack is still under investigation \cite{GMCS_security}.

\section{Practical issues}

In this section, we will discuss several practical issues in
implementing the ``dual detector'' idea, including the loss
introduced by the optical switch, the probability of using each type
of detectors, the chromatic dispersion of long fiber and the
security of a practical setup.

In previous sections, we assume that Bob has an ideal, lossless
optical switch to distribute the incoming pulses between the two
detectors. A commercial high speed optical switch designed for
telecom industry has a insertion loss around 3dB. To make a fair
comparison, we introduce an additional 3dB loss in Bob's system for
``dual detector'' scheme. The simulation results demonstrate that in
the case of single photon QKD, the advantage of ``dual detector'' is
still obvious, as shown in Fig.8 and Fig.9, while in the case of
decoy state QKD and GMCS QKD, the additional 3dB loss is disastrous:
with the parameters used in Sections 3 and 4, the ``dual detector''
scheme shows no advantage over conventional ``single detector''
scheme. This result is not surprising: for decoy-state QKD, even
with a perfect lossless switch, the improvement is quite limited
(see Fig.4); for GMCS QKD, we already know that the key rate drops
sharply as the channel loss increase \cite{nature2003}.

We remark that the 3dB loss of a commercial high speed optical
switch is mostly due to the fiber-waveguide coupling loss, which is
by no means a hard limit imposed by the technology. In fact, if only
one wavelength channel is used for QKD, one could optimize waveguide
design to minimize coupling loss. In this case, one can reasonably
expect the insertion loss to be much lower than 1dB, at a higher
price.

\begin{figure}[!t]
\resizebox{8.5cm}{!}{\includegraphics{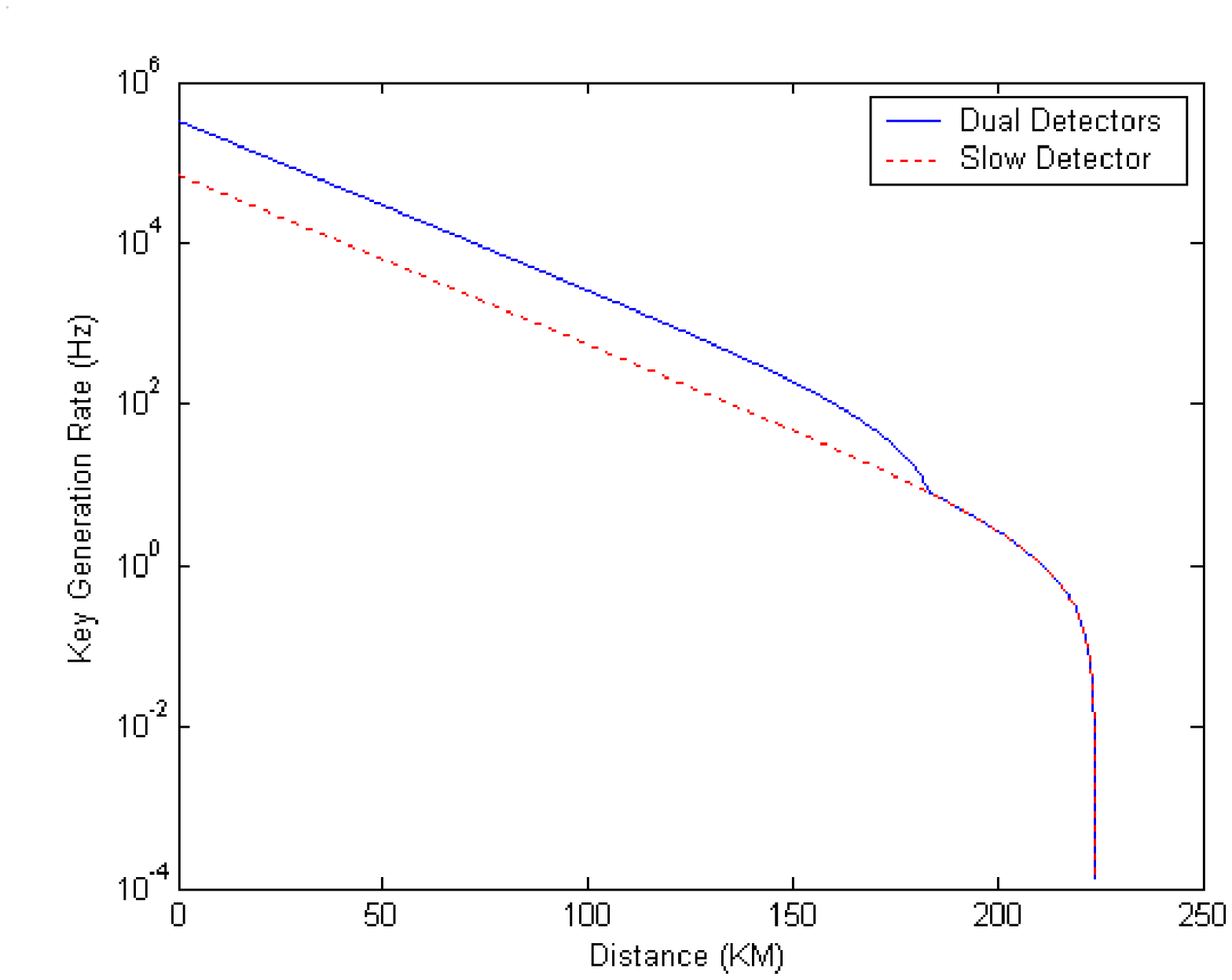}}
\caption{Simulation results with a lossy (3dB) optical switch.
Simulation parameters: $r_1=10\mathrm{GHz}$,
$\eta_{\mathrm{D}}^{(1)}=0.0027$, $Y_{0}^{(1)}=3.2\times10^{-9}$ and
$e_{\mathrm{det}}^{(1)}=0.097$ \cite{SPD_10G2}. Other parameters are
same as in Fig.1. The key rate of ``dual detectors'' system is
significantly higher than either of the two single SPD systems up to
$\sim180\mathrm{km}$. Note that no secure key can be produced by
SPD1 alone at any distance.}\label{fig8}
\end{figure}

\begin{figure}[!t]
\resizebox{8.5cm}{!}{\includegraphics{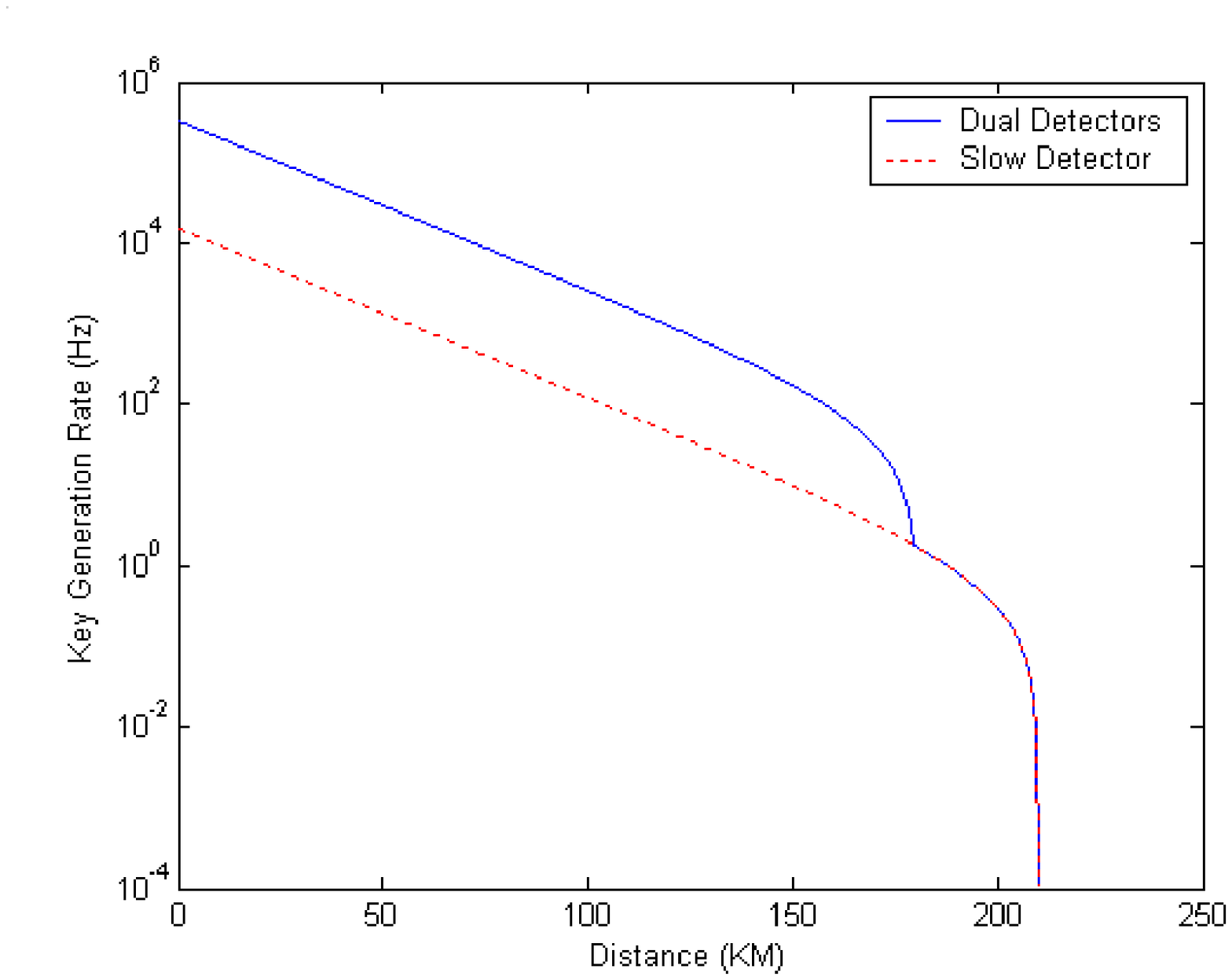}}
\caption{Simulation results with a lossy (3dB) optical switch.
Simulation parameters: $r_1=10\mathrm{GHz}$,
$\eta_{\mathrm{D}}^{(1)}=0.0027$, $Y_{0}^{(1)}=3.2\times10^{-9}$ and
$e_{\mathrm{det}}^{(1)}=0.097$ \cite{SPD_10G2};
$r_2=100\mathrm{MHz}$, $\eta_{\mathrm{D}}^{(2)}=0.0027$,
$Y_{0}^{(2)}=3.2\times10^{-9}$ and $e_{\mathrm{det}}^{(2)}=0.018$;
Other parameters are same as in Fig.1. The key rate of ``dual
detectors'' system is significantly higher than either of the two
single SPD systems up to $\sim175\mathrm{km}$. Note that no secure
key can be produced by SPD1 alone at any distance.}\label{fig9}
\end{figure}

Another important issue is how to determine the probability of using
each of the two detectors. Since only the output from the fast
detector contributes to the final key, in asymptotic limit, the
probability of using the slow detector should be as small as
possible. In practice, two other factors have to be taken into
account. First, in order to estimate the system parameters
accurately, Alice/Bob have to acquire enough data for either type of
detectors in a reasonable time period. This determines the lower
bound on the probability for choosing the slow detector. Second, the
slow detector may have a large time jitter. If more than one pulses
are sent to it within its response window, Bob cannot tell which
incoming pulse the detection event corresponds to and the QBER will
increase. This determines the upper bound on the probability of
choosing the slow detector. In the following, we will estimate the
probability $p$ of choosing the slow detector (detector2) based on
the parameters of a practical setup.

We assume the period of the signal pulse is $T_{\mathrm{sig}}$, and
the time resolution (time jitter) of detector2 is
$T_{\mathrm{det}}$. In each single response window of detector2,
there are $k$ ($=T_{\mathrm{det}}/T_{\mathrm{sig}}$) pulses sent out
by Alice. Bob randomly chooses to use either detector1 (with a
probability of $1-p$) or detector2 (with a probability of $p$) to
measure the input pulse. In each $T_{\mathrm{det}}$ time window, the
probabilities that Bob does not choose detector2, chooses it one
time, or chooses it more than one time are $P_{0}$ ($=(1-p)^{k}$),
$P_{1}$ ($=kp(1-p)^{k-1}$) and $P_{M}$ ($=1-P_{0}-P_{1}$),
respectively. Assuming that $p\ll{k}\ll{1}$, we have $P_{1}
=kp-k(k-1)p^{2}$ and $P_{M}=k(k-1)p^{2}/2$. Note that the
probability that Bob chooses detector2 only one time (in the
$T_{\mathrm{det}}$ time window)and he does detect a signal is
$P_{\mathrm{sig}}=\mu\eta P_{1}$, where $\mu$ is the average photon
number per pulse, and $\eta$ is the overall transmission efficiency
(including the channel efficiency, the optical transmittance in
Bob's system, and the efficiency of detector2). This is an effective
detection. On the other hand, if Bob chooses detector2 more than one
time and he does detect a signal, then he has to randomly assign
this detection event to one of the input pulses he chooses. If we
assume that the major contribution to $P_{M}$ comes from $P_{2}$,
then the probability for Bob to get a ``messed detection'' is
$P_{\mathrm{err}}=2\mu\eta P_{M}$, where the factor $2$ takes into
account that two pulses have been sent to detector2. The error rate
of these ``messed detection'' is $1/4$, because half of the time,
Bob will assign the detection event to the right pulse (no error),
the other half of time, Bob will assign the detection event to the
wrong pulse ($1/2$ error). The overall QBER due to the
``multi-pulses'' problem can be estimated as
\begin{equation}
QBER\approx\frac{P_{\mathrm{err}}}{4(P_{\mathrm{err}}+P_{\mathrm{sig}})}\approx
\frac{1}{4}(k-1)p
\end{equation}

Using parameters in Fig.1, $T_{\mathrm{det}}=100ns$ \cite{TES2},
$T_{\mathrm{sig}}=1ns$ ($1\mathrm{GHz}$ pulse repetition rate), we
have $k=100$. To make the additional $QBER<1\%$, we get
$p<4\times10^{-4}$. On the other hand, if we assume the channel loss
is 21dB (100km fiber), $G_{\mathrm{Bob}}=0.16$,
$\eta_{\mathrm{D}}^{(2)}=0.5$, the additional loss due to optical
switch is 3dB, then, with $p=4\times10^{-4}$ and $1\mathrm{GHz}$
pulse repetition rate, Bob will have $\sim10^{6}$ counts in about 2
hours, which is large enough to estimate various parameters of the
QKD system \cite{stat_flu}. In Fig.3, since both detectors have
small time jitter, the $p$ value can be relatively large.

We remark that the minimum $p$ achievable in practice is limited by
the extinction ratio of the optical switch. On the other hand, it
may be possible to overcome this ``multi pulses'' problem by
improving the protocol. For example, Bob can prepare his random
pattern for the optical switch in the following way: if the $n_{th}$
pulse is assigned to the slow detector, then the next $r$ pulses
($r$ is determined by the time resolution of the slow detector) will
not be assigned to it. This is equivalent to introducing a ``virtual
dead time'' to the slow detector. It is interesting to investigate
the security of this scheme. However, we do not have a definite
answer so far.

We remark that the slow response of detector2 also prevents Bob from
using a passive beam splitter to replace the optical switch. In that
case, Bob cannot tell which input pulse corresponds to the
detection event from detetor2.

The third practical issue is the chromatic dispersion introduced by
the telecom fiber. The chromatic dispersion of conventional telecom
fiber at $1550\mathrm{nm}$ is around
$18\mathrm{ps/nm}\cdot{\mathrm{km}}$. In many QKD system, the
spectral width of the laser pulse is in the order of
$0.1\mathrm{nm}$, so the temporal pulse width will be extended by
$180\mathrm{ps}$ after it goes through $100\mathrm{km}$ fiber. This
will cause severe cross talks between adjacent pulses when the
system is operated at $10\mathrm{GHz}$. We remark that dispersion
compensation (DC) is an important issue even in classical
communication, and various successful DC techniques have been
developed. For example, in \cite{dispersion}, after going through a
50km fiber, a 460fs pulse was only slightly broadened to 470fs.
Similar techniques can also be applied to a QKD system. We remark
that the loss introduced by DC components will not compromise the
performance of the QKD system, since it can be deployed inside
Alice's system.

An important assumption of our ``dual detector'' idea is that a
signal from Eve cannot fool the two detectors by behaving
differently. Such an assumption must not be taken for granted.
Instead, it should be examined carefully in any practical system.
However, we note that there are various defense strategies that
Alice and Bob can employ to make our assumption more realistic. For
instance, to prevent Eve from attacking the two detectors
differently by sending laser pulses at different wavelengths, Bob
has to make sure that the spectral responses of the two detectors
are identical to Eve. Normally, a photon detector has a spectral
response range from tens of nm to larger than 100nm, while the
spectral width of the laser pulse from Alice is less than 1nm. By
placing a narrowband optical filter (with a bandwidth of
$\sim1\mathrm{nm}$) at the entrance of Bob's system, we can safely
assume that the spectral responses of both detectors are flat in
this spectral window \cite{filter}. On the other hand, Eve may
explore the different temporal responses of the two detectors by
shifting the arriving time of the laser pulse \cite{time-shift}. For
example, in the case of up-conversion SPD, to achieve a low dark
count, Bob uses narrow time windows, which are centered around the
incoming pulses, to post-select effective detection events. All
detection events outside these time windows will be dropped. If the
widths of time windows are different for the two detectors, Eve may
time-shift a pulse in such a way that one detector will treat it as
an effective event, while the other one will drop it. We remark that
to prevent Eve from launching such a time-shift attack, Bob should
monitor the time distribution of all his detection events.

\section{Discussion}

The performance of a QKD system in telecom wavelength is mainly
determined by the performance of its detection system. To achieve
high speed, long distance QKD, fast and quiet detectors are on
demand. Unfortunately, in practice, a fast detector is usually more
noisy than a slow one. Here, we propose a ``dual detectors'' scheme
to improve the performance of a practical QKD system with realistic
detectors. Our simulation results demonstrate significant
improvements of the secure key rate in some QKD protocols.

Any security proof of a practical QKD system is built on its
underlying assumptions: what kinds of imperfections exist, what Eve
can control/know about Alice's and Bob's systems. Obviously, if we
allow Eve to control/know everything (like which SPD clicks in BB84
QKD), secure QKD is hopeless. On the other hand, people normally
assume that the loss inside Bob's system and the dark count of Bob's
SPD are under Eve's control. In this case, secure QKD is still
possible. Unfortunately, in practice, there are no clear rules to
determine what assumptions should be chosen. Some assumptions may
enforce the security of a QKD system without comprising its
efficiency, while others may damage its efficiency greatly without
contributing much to its security. It is important to inspect all
those underlying assumptions behind a practical QKD system
carefully. It will be very interesting to test experimentally our
assumption---that a signal cannot fool the two detectors by behaving
differently---in a practical QKD system. Such a test will lead to a
better understanding and potential refinements of our assumption.

Financial support from NSERC, CIAR, CRC Program, CFI, OIT, MITACS, PREA and
CIPI are gratefully acknowledged. This research was supported by
Perimeter Institute for Theoretical Physics. Research at Perimeter
Institute is supported in part by the Government of Canada through
NSERC and by the province of Ontario through MEDT.


\begin{thebibliography}{00}
\bibitem{Gisin} N. Gisin suggested that we should be less generous to Eve in the sense that we can assume that Eve cannot change the dark count of a detector.
\bibitem{Dark_count}
We remark that in this paper, we assume that the background noise of
Bob's detector is dominated by its intrinsic noise. This assumption
may be not applicable in some QKD setups, where strong
synchronization pulses and quantum signals go through the same
fiber. In that case, the background noise may be dominated by the
stray light from the intense synchronization pulses.
\bibitem{stat_flu}
For example, if $QBER=1\%$, among the total $10^6$ detections, the
error counts is $\sim10^4$. This gives us an estmation of the QBER
as $(1\pm0.01)\%$, which is accurate enough in practice.
\bibitem{filter}
As pointed out by an anonymous reviewer, in the case of
up-conversion detectors, the acceptance bandwidth of the nonlinear
process defines the spectral filtering. This is already less than
$1nm$. Thus, no additional bandpass filter is required.
\end{thebibliography}
\end{document}